\documentclass[fp]{jpsj3}
\usepackage{txfonts}
\usepackage{bm}
\usepackage{physics}
\usepackage{comment}
\usepackage[dvipdfmx]{color}
\usepackage[dvipdfmx]{hyperref}
\makeatletter
\providecommand{\ext@figure}{lof}
\providecommand{\ext@table}{lot}
\makeatother

\title{Nonlocal Cooper pairs in finite topological superconductors\\
and their relation to Majorana nonlocality}

\author{Hiroto Mizoguchi$^1$, Yutaro Nagae$^1$, Yasuhiro Asano$^2$, and Satoshi Ikegaya$^2$}
\inst{$^1$ Department of Applied Physics, Nagoya University, Nagoya 464-8603, Japan\\
$^2$ Department of Applied Physics, Hokkaido University, Sapporo 060-8628, Japan}

\abst{We identify two fundamental properties of the Gor'kov Green's function of finite one-dimensional topological superconductors.
In the low-frequency (low-energy) regime, the normal and anomalous Green's functions become identical up to a phase factor.
Moreover, they exhibit pronounced nonlocality:
the components connecting the two ends of the system grow exponentially with system length, whereas the local components at either end vanish in the zero-frequency limit.
These striking features signify the emergence of unconventional nonlocal Cooper pairs
associated with a nonlocal fermionic mode composed of hybridized Majorana end modes.
The nonlocal Cooper pairs are directly linked to fermion parity and to the nonlocal transport properties of finite topological superconductors.
By uncovering the unusual nonlocal properties of the Gor'kov Green's function,
our analysis advances the understanding of Majorana nonlocality, a key concept in topological quantum computation.}

\begin{document}
\maketitle

\section{Introduction}
%****************************************************************************************************************
In the mean-field theory of superconductivity, two complementary perspectives can be adopted.
One focuses on Bogoliubov quasiparticle excitations described by the Bogoliubov--de Gennes (BdG) equations,
while the other focuses on Cooper-pair correlations described by the Gor'kov equations.
For example, the superconducting proximity effect can be interpreted
in the quasiparticle picture as the formation of correlated electron--hole pairs via Andreev reflection at junction interfaces~\cite{andreev_64,beenakker_97,courtois_00}.
In contrast, in the pair-correlation picture, it corresponds to the penetration of Cooper pairs into an adjacent material~\cite{gennes_64,eilenberger_68,usadel_70}.
The complementary use of these two viewpoints is essential for a comprehensive understanding of superconducting phenomena.

Over the past two decades, Majorana end modes in topological superconductors have been extensively studied within the quasiparticle framework%
~\cite{green_00,kitaev_01,kane_10,zhang_11,sato_16,strinati_26}.
In parallel, odd-frequency Cooper pairs have attracted considerable attention from the perspective of pair correlations~\cite{berezinskii_74,efetov_05,tanaka_07,linder_19}.
It is now widely recognized that these two concepts are intimately connected:
odd-frequency pairing emerges wherever Majorana end modes exist~\cite{nagaosa_12,asano_13,cayao_20},
as a consequence of bulk--boundary correspondence~\cite{gurarie_11,tamura_19,daido_19}.
The Majorana quasiparticle perspective has proven particularly useful for understanding unusual quantum transport phenomena%
~\cite{tanaka_04,tanaka_05(1),law_09,ikegaya_15,ikegaya_16(1)},
whereas the odd-frequency pairing perspective has often been employed to interpret unconventional paramagnetic magnetic responses~\cite{asano_11,higashitani_14,asano_14,lutchyn_17}.
Thus, these complementary viewpoints have also played a pivotal role in the development of the physics of topological superconductivity.
However, the connection between Majorana quasiparticles and unconventional Cooper pairs has been established only for \emph{semi-infinite} one-dimensional topological superconductors.

In \emph{finite} one-dimensional topological superconductors,
Majorana end modes localized at opposite ends hybridize to form a single nonlocal fermionic mode.
The quantum nonlocality associated with this mode---referred to as Majorana nonlocality---has been a central issue in the field of topological superconductivity,
because the occupation of this nonlocal mode encodes a qubit for topological quantum computation~\cite{ivanov_01,kitaev_03,sarma_08,fisher_11,alicea_16}.
A variety of theoretical proposals have been put forward to detect Majorana nonlocality through multiterminal transport measurements%
~\cite{demler_07,tewari_08,beenakker_08,fu_10,egger_11,hu_13,law_13,rosenow_13,tewari_15,flensberg_20,xu_20,fujimoto_23,ikegaya_24}.
Most analyses, however, rely on a simplified low-energy effective Hamiltonian~\cite{kitaev_01},
$H_M=i(E_M/2)\gamma_a \gamma_b$, where $\gamma_a=\gamma_a^{\dagger}$ and $\gamma_b=\gamma_b^{\dagger}$
represent Majorana end modes localized at opposite ends, and $E_M$ denotes the coupling energy~\cite{stanescu_12}.
Consequently, Majorana nonlocality has been investigated predominantly within the quasiparticle framework,
while its manifestation in terms of pair correlation perspective remains largely unexplored.
It has been already pointed out that, at special parameter values, including the Kitaev sweet spot~\cite{kitaev_01},
a fixed-parity end-to-end correlation function can remain finite even in the thermodynamic limit~\cite{zhou_18}.
In contrast, the properties of thermal-equilibrium Green's functions in finite topological superconductors under more general conditions have not yet been analyzed in detail~\cite{tanaka_24}.
In this work, we address this issue.

We analytically derive the Gor'kov Green's function for finite one-dimensional topological superconductors.
In the low-frequency regime, the Green's function exhibits two distinctive properties.
First, the normal and anomalous Green's functions%
---associated with single-particle and Cooper-pair correlations, respectively---are equivalent up to a phase factor.
Second, they display pronounced nonlocality: the end-to-end components of the Green's function grow exponentially with the system length,
whereas the local components at either end vanish in the zero-frequency limit.
These striking features signify the emergence of unconventional nonlocal Cooper pairs in finite topological superconductors.
Such nonlocal Cooper pairs are directly linked to fermion parity and to nonlocal transport phenomena previously investigated within the quasiparticle framework.
Our results provide a foundation for revisiting Majorana nonlocality from the perspective of pair correlations.

%****************************************************************************************************************
\section{Gor'kov Green's function}
%****************************************************************************************************************
%----------------------------------------------------------------------------
\subsection{Derivation of Green's function}
%----------------------------------------------------------------------------
In this section, we derive the Gor'kov Green's function for a one-dimensional topological superconductor.
Specifically, we employ the Kitaev chain model~\cite{kitaev_01},
which captures the essential low-energy physics of a broad class of topological superconductors,
including superconductor--semiconductor hybrid nanowires~\cite{sarma_10,oreg_10,kouwenhoven_12,deng_12},
magnetic atom chains on superconducting substrates~\cite{beenakker_11,yazdani_13,yazdani_14,yazdani_17},
and planar topological Josephson junctions~\cite{flensberg_17,halperin_17,nichele_19,yacoby_19,ikegaya_22}.
Our mean-field theory is formulated within the grand-canonical framework and does not impose strict particle-number conservation,
in contrast to the particle-number-conserving formulation employed in Ref.~\cite{zhou_18}.
We note that the artificial topological superconducting systems discussed above incorporate superconducting electrodes.
Consequently, their low-energy effective descriptions based on the Kitaev chain naturally adopt the BdG formalism, in which strict particle-number conservation is not imposed.
The BdG Hamiltonian is given by~\cite{kitaev_01}
\begin{align}
H = \sum_j \Big[
-t \left(c_{j+1}^{\dagger} c_j + \mathrm{h.c.}\right)
- \mu \left(c_j^{\dagger} c_j - \tfrac{1}{2} \right) +
 \left( \frac{i\Delta}{2} e^{i\theta}
c_{j+1}^{\dagger} c_j^{\dagger} + \mathrm{h.c.} \right)
\Big],
\label{eq:original_bdg}
\end{align}
where $c_j$ ($c_j^\dagger$) annihilates (creates) an electron at site $j$,
$t$ is the hopping amplitude, and $\mu$ is the chemical potential.
The pair potential has amplitude $\Delta$ and phase $\theta$.
In what follows, we set $\theta = 0$ for simplicity and take the lattice constant to be unity.
Introducing the Nambu basis, the Hamiltonian can be written as
\begin{align}
\begin{split}
&H = \frac{1}{2} \sum_{j,j'}  \left(c_{j}^{\dagger}, c_{j}\right) \hat{H}(j,j')
\begin{pmatrix} c_{j'}\\ c_{j'}^{\dagger} \end{pmatrix},\\
&\hat{H}(j,j') = \begin{cases}
\hat{T} & \text{for } j'=j-1  \\
\hat{T}^{\dagger} & \text{for } j'=j+1  \\
\hat{K} & \text{for } j'=j  \\
0 & \text{otherwise}
\end{cases},\quad
\hat{T}= -t \hat{\tau}_3 + i\frac{\Delta}{2} \hat{\tau}_1, \quad
\hat{K} = -\mu \hat{\tau}_3,
\end{split}
\label{eq:matrix_bdg}
\end{align}
where $\hat{\tau}_{1,2,3}$ denote the Pauli matrices in Nambu space.

The Matsubara Gor'kov Green's function $\hat{\mathcal{G}}_0(j,j')$ for the bulk system satisfies the Gor'kov equation
\begin{align}
\sum_{j_1} \left[i\omega \delta_{j,j_1}- \hat{H}(j,j_1)\right] \hat{\mathcal{G}}_0(j_1,j') = \delta_{j,j'},
\label{eq:bulk_gorkov}
\end{align}
where $\omega=(2n+1)\pi T$ is the fermionic Matsubara frequency at temperature $T$.
In the regime $\Delta,\omega \ll t$,
 the Green's function takes the form
\begin{align}
\hat{\mathcal{G}}_0(j,j')
= -\frac{i}{v_F \Omega_{k_F}}
\left(
a_0 + a_1 \hat{\tau}_1 + a_3 \hat{\tau}_3
\right),
\label{eq:bulk_green}
\end{align}\\
where
\begin{align}
\begin{split}
&a_0=\omega \cos\!\big(k_F |j-j'|\big) e^{-|j-j'|/\xi},\\
&a_1=\Delta_{k_F} \sin\!\big(k_F (j-j')\big) e^{-|j-j'|/\xi},\\
&a_3=i\Omega_{k_F}\sin \!\big(k_F |j-j'|\big) e^{-|j-j'|/\xi},\\
&\Delta_k = \Delta \sin k, \quad
\Omega_k=\sqrt{\omega^2 + \Delta_k^2}, \\
&k_F = \arccos \! \left( -\frac{\mu}{2t} \right), \quad
v_F = 2t \sin k_F,\quad
\xi = \frac{v_F}{\Omega_{k_F}}.
\end{split}
\end{align}
A detailed derivation of Eq.~(\ref{eq:bulk_green}) is provided in Appendix~\ref{sec:app_bulk}.

To obtain the Green's function for a finite system, we introduce boundary potentials at $j=0$ and $j=L+1$ and solve the Dyson equation
\begin{align}
\hat{\mathcal{G}}_V(j,j')
= \hat{\mathcal{G}}_0(j,j') 
+\sum_{j_1,j_2}
\hat{\mathcal{G}}_0(j,j_1)\hat{V}(j_1,j_2)\hat{\mathcal{G}}_V(j_2,j'),
\end{align}
with
\begin{align}
\hat{V}(j,j')
= V\,\delta_{j,j'} \bigl(\delta_{j,0}+\delta_{j,L+1}\bigr) \hat{\tau}_3.
\end{align}
Taking the hard-wall limit, we obtain the Green's function of the finite topological superconductor for $1\leq j,j' \leq L$:
\begin{align}
\hat{\mathcal{G}}(j,j') = \lim_{V\to\infty} \hat{\mathcal{G}}_V(j,j').
\end{align}
The explicit expression of $\hat{\mathcal{G}}(j,j')$ is given in Eq.~(\ref{eq:green_finite_full}) of Appendix~\ref{sec:app_finite}.
We note that an equivalent Green's function for a continuum model of a $p$-wave superconductor was derived using McMillan's method in Ref.~[\citen{tanaka_24}];
however, a detailed analysis of its properties has not yet been carried out.
For later convenience, we decompose the Green's function in Nambu space as
\begin{align}
\hat{\mathcal{G}}(j,j')=\begin{pmatrix} \mathcal{G}(j,j') & \mathcal{F}(j,j') \\ \underline{\mathcal{F}}(j,j') & \underline{\mathcal{G}}(j,j') \end{pmatrix},
\end{align}
where $\mathcal{G}$ ($\underline{\mathcal{G}}$) denotes the normal Green's function for electrons (holes),
while $\mathcal{F}$ and $\underline{\mathcal{F}}$ denote the anomalous Green's functions.

%----------------------------------------------------------------------------
\subsection{Fundamental properties}\label{sec:fundamental_properties}
%----------------------------------------------------------------------------
In this section, we analyze the Green's functions in the low-frequency regime,
$\omega \ll \Delta_{k_F}$, for a system of length $L$ satisfying
$1 \ll \xi_0 \ll L$, where $\xi_0=v_F/\Delta_{k_F}$.
Under these conditions, we retain only the leading-order contributions in
\begin{align}
\delta = \frac{\omega}{\Delta_{k_F}},\;
\frac{1}{\xi_0},\;
e^{-L/\xi_0}.
\end{align}
Accordingly, the Green's functions reduce to

\begin{align}
\begin{split}
\hat{\mathcal{G}}(1,j')=&-i \frac{\omega}{\omega^2 + E_M^2} \frac{2}{\xi_0} \sin (k_F) \sin(k_F j')
\left(1 - \hat{\tau}_2\right) e^{-j'/\xi_0}\\
&-i \frac{E_M}{\omega^2 + E_M^2} \frac{2}{\xi_0} \sin (k_F) \sin(k_F (L+1-j'))
\left(-\hat{\tau}_1 - i\hat{\tau}_3\right) e^{-(L-j')/\xi_0},
\label{eq:green_low(1)}
\end{split}
\end{align}
\begin{align}
\begin{split}
\hat{\mathcal{G}}(j,L)=&-i \frac{\omega}{\omega^2 + E_M^2}  \frac{2}{\xi_0} \sin(k_F (L+1-j)) \sin (k_F)
\left(1 + \hat{\tau}_2\right) e^{-(L-j)/\xi_0}\\
&-i \frac{E_M}{\omega^2 + E_M^2}\frac{2}{\xi_0} \sin (k_F j) \sin(k_F)
\left(\hat{\tau}_1 - i\hat{\tau}_3\right) e^{-j/\xi_0}.
\label{eq:green_low(2)}
\end{split}
\end{align}

where
\begin{align}
E_M = 2 \Delta_{k_F} \sin\!\big(k_F (L+1)\big) e^{-L/\xi_0}.
\end{align}
Detailed derivations of Eqs.~(\ref{eq:green_low(1)})  and (\ref{eq:green_low(2)}) are provided in Appendix~\ref{sec:app_finite}.
Importantly, the second term in Eq.~(\ref{eq:green_low(1)}) [Eq.~(\ref{eq:green_low(2)})] \emph{increases} exponentially with increasing $j'$ [decreasing $j$],
whereas the first term decays exponentially with increasing $j'$ [decreasing $j$].
Both terms are suppressed near the center of the topological superconductor, i.e., $j' \sim L/2$ [$j \sim L/2$].
To elucidate the properties of the Green's functions, we focus on those associated with the outermost sites:
\begin{align}
\begin{split}
\hat{\mathcal{G}}(1,1)
= -i C \frac{\omega}{\omega^2 + E_M^2}
\left(1 - \hat{\tau}_2\right), \qquad
\hat{\mathcal{G}}(L,L)
= -i C \frac{\omega}{\omega^2 + E_M^2}
\left(1 + \hat{\tau}_2\right),
\label{eq:local_green}
\end{split}
\end{align}
and
\begin{align}
\begin{split}
\hat{\mathcal{G}}(1,L)
= -i C \frac{E_M}{\omega^2 + E_M^2}
\left(-\hat{\tau}_1 - i\hat{\tau}_3\right), \qquad
\hat{\mathcal{G}}(L,1)
= -i C \frac{E_M}{\omega^2 + E_M^2}
\left(\hat{\tau}_1 - i\hat{\tau}_3\right),
\label{eq:nonlocal_green}
\end{split}
\end{align}
where
\begin{align}
&C = \frac{2}{\xi_0} \sin^2 k_F.
\end{align}
We now turn to the two key features of these Green's functions.

First, for both the local and nonlocal components, the normal and anomalous Green's functions differ only by a phase factor:
\begin{align}
\mathcal{G}(j,j')=\pm i e^{i \theta} \mathcal{F}(j,j'),
\label{eq:prop1}
\end{align}
where the superconducting phase $\theta$ is included explicitly.
Recalling the relations between the Green's functions and the field operators,
\begin{align}
\begin{split}
\langle c_j c_{j'}^{\dagger}\rangle = -T\sum_{\omega} \mathcal{G}(j,j'),\qquad
\langle c_j c_{j'}\rangle = -T\sum_{\omega} \mathcal{F}(j,j'),
\end{split}
\end{align}
we find that Eq.~(\ref{eq:prop1}) implies that electrons and holes are equivalent up to a phase factor.
Thus, this equivalence between the normal and anomalous Green's functions reflects the Majorana character of the low-energy physics.
We note that a similar property of the \emph{local} Green's functions in \emph{semi-infinite} topological superconductors has been investigated previously~\cite{asano_13}.

Second, the Green's functions exhibit pronounced nonlocality.
As shown in Eq.~(\ref{eq:nonlocal_green}), the nonlocal Green's functions, $\hat{\mathcal{G}}(1,L)$ and $\hat{\mathcal{G}}(L,1)$, are proportional to
\begin{align}
\frac{E_M}{\omega^2 + E_M^2}, \nonumber
\end{align}
which represents a Lorentzian peak centered at zero frequency, with a width set by $E_M \propto e^{-L/\xi_0}$.
Accordingly, the nonlocal Green's functions retain significant amplitudes for $\omega \ll E_M$.
In particular, at zero frequency, we obtain
\begin{align}
\begin{split}
\lim_{\omega\rightarrow0}\hat{\mathcal{G}}(1,L) \propto e^{L/\xi_0}, \qquad
\lim_{\omega\rightarrow0}\hat{\mathcal{G}}(L,1) \propto e^{L/\xi_0},
\label{eq:prop2}
\end{split}
\end{align}
indicating that the nonlocal Green's functions \emph{grow} exponentially with the system length.
In contrast, as shown in Eq.~(\ref{eq:local_green}), the local Green's functions, $\hat{\mathcal{G}}(1,1)$ and $\hat{\mathcal{G}}(L,L)$, are proportional to
\begin{align}
\frac{\omega}{\omega^2 + E_M^2}, \nonumber
\end{align}
exhibiting odd-frequency symmetry and therefore vanishing at zero frequency:
\begin{align}
\begin{split}
\lim_{\omega\rightarrow0}\hat{\mathcal{G}}(1,1)=0,\qquad
\lim_{\omega\rightarrow0}\hat{\mathcal{G}}(L,L)=0.
\label{eq:prop2-2}
\end{split}
\end{align}
Consequently, the low-energy physics in the regime $\omega \ll E_M$ is dominated by nonlocal Green's functions.
We note that in typical superconductors, nonlocal Green's functions usually decay exponentially over a length scale $\xi_0$ [see also Eq.~(\ref{eq:bulk_green})].
Thus, the present results clearly demonstrate the striking and intrinsic nonlocal character of the Green's functions in finite topological superconductors.

\begin{figure}[t]
\begin{center}
\includegraphics[width=0.9\textwidth]{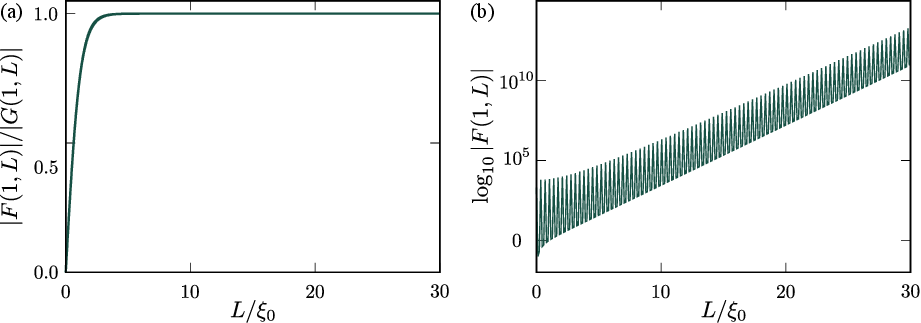}
\caption{(a) $|\mathcal{F}(1,L)|/|\mathcal{G}(1,L)|$ at $\omega = 0$ as a function of the system length $L$.
(b) $\log_{10} |\mathcal{F}(1,L)|$ at $\omega = 0$ as a function of $L$.}
\label{fig:figure1}
\end{center}
\end{figure}

Overall, we obtain anomalous Green's functions satisfying
\begin{align}
|\mathcal{F}(1,L)| = |\mathcal{G}(1,L)|,
\end{align}
together with
\begin{align}
\mathcal{F}(1,L) \propto e^{L/\xi_0}.
\end{align}
From the viewpoint of pair correlations, these properties signify the emergence of unconventional nonlocal Cooper pairs in finite topological superconductors.
To verify the robustness of these analytical findings, we present numerical results obtained without approximation,
where the Green's functions are computed using the recursive Green's function method~\cite{fisher_81}.
In Fig.~\ref{fig:figure1}(a), we plot $|\mathcal{F}(1,L)|/|\mathcal{G}(1,L)|$ at $\omega = 0$ as a function of the system length $L$.
The parameters are set to $\mu=-t$ and $\Delta=0.001t$.
Over a wide range of $L$, we find $|\mathcal{F}(1,L)| / |\mathcal{G}(1,L)| = 1$,
demonstrating the equivalence of the normal and anomalous Green's functions, consistent with Eq.~(\ref{eq:prop1}).
Figure~\ref{fig:figure1}(b) shows $\log_{10} |\mathcal{F}(1,L)|$ at $\omega = 0$ as a function of $L$.
We clearly observe the exponential growth of the nonlocal Green's function predicted in Eq.~(\ref{eq:prop2}).
Since $E_M \propto \sin(k_F L)$, $\mathcal{F}(1,L)$ also exhibits rapid oscillations as a function of $L$.
Although our analytical results are derived under the assumption, $1 \ll \xi_0 \ll L$,
we confirm that the key properties in Eqs.~(\ref{eq:prop1}) and (\ref{eq:prop2}) remain valid over a broad range of $L$.
%----------------------------------------------------------------------------
%----------------------------------------------------------------------------

%----------------------------------------------------------------------------
\subsection{Relation to fermion parity}
%----------------------------------------------------------------------------

In this section, we examine the relationship between the nonlocal Green's functions and the emergent nonlocal fermionic mode originating from Majorana end modes.
We note that previous studies have established that the end-to-end correlation function within a particle-number-conserving framework
is directly related to the fermion parity~\cite{sato_16,zhou_18}.
Extending this result, we demonstrate that the anomalous Green's function in thermal equilibrium likewise exhibits a direct relation to the fermion parity.
As a starting point, we consider the end-to-end correlator in thermal equilibrium
\begin{align}
\langle c_1 c_{L}^{\dagger}\rangle = -T\sum_{\omega} \mathcal{G}(1,L).
\end{align}
For sufficiently low temperatures (i.e., $T \ll \Delta$) and for system lengths satisfying Eq.~(\ref{eq:approx2}), this correlator reduces to
\begin{align}
\begin{split}
\langle c_1 c_{L}^{\dagger} \rangle &\sim CT \sum_{\omega} \frac{E_M}{\omega^2 + E_M^2}=\frac{C}{2} \left\{1-2f(E_M) \right\},
\end{split}
\end{align}
where $f(E)$ denotes the Fermi distribution function.
Contributions from the continuum modes to the nonlocal correlator are exponentially suppressed and may therefore be neglected.
We next consider the low-energy effective Hamiltonian of the topological superconductor~\cite{kitaev_01},
\begin{align}
H_M=i\frac{E_M}{2}\gamma_a \gamma_b,
\end{align}
where $\gamma_a$ and $\gamma_b$ represent Majorana end modes localized at $j=1$ and $j=L$, respectively.
Introducing the nonlocal fermionic mode
\begin{align}
f=\frac{\gamma_a + i \gamma_b}{2},
\end{align}
the effective Hamiltonian can be written in diagonal form as~\cite{kitaev_01}
\begin{align}
H_M = E_M \left( n_f-\frac{1}{2} \right), \quad n_f = f^{\dagger}f.
\end{align}
In thermal equilibrium, this yields $\langle n_f \rangle = f(E_M)$.
Using this relation, the nonlocal correlator becomes
\begin{align}
\langle c_1 c_{L}^{\dagger} \rangle &\sim \frac{C}{2} \left(1-2\langle n_f \rangle\right) = \frac{C}{2} \langle P_f \rangle,
\end{align}
where
\begin{align}
P_f = -i\gamma_a \gamma_b = (-1)^{n_f},
\end{align}
denotes the fermion-parity operator associated with the occupation of the nonlocal fermionic mode~\cite{kitaev_01}.
We therefore arrive at
\begin{align}
\langle P_f \rangle=-\frac{2T}{C}\sum_{\omega} \mathcal{G}(1,L).
\end{align}
On the basis of the equivalence in Eq.~(\ref{eq:prop1}), the anomalous Green's function is likewise directly related to the fermion parity:
\begin{align}
\langle P_f \rangle=-\frac{2iT}{C}\sum_{\omega} \mathcal{F}(1,L),
\end{align}
which demonstrates a direct connection between unconventional nonlocal Cooper pairs and the fermion parity of the topological superconductor.

%----------------------------------------------------------------------------
\subsection{Role in charge transport}
\begin{figure}[t]
\begin{center}
\includegraphics[width=0.5\textwidth]{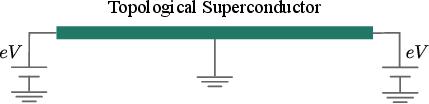}
\caption{Schematic illustration of a multiterminal setup with normal leads attached to both ends of a topological superconductor.}
\label{fig:figure2}
\end{center}
\end{figure}

%----------------------------------------------------------------------------
In this section, we examine the role of the nonlocal Green's functions in charge transport.
To this end, we consider the multiterminal setup shown in Fig.~\ref{fig:figure2}, where normal leads are attached to both ends of a topological superconductor.
A same bias voltage is applied to the normal leads, while the topological superconductor is grounded.
Charge transport in this system has previously been investigated within the quasiparticle framework.
We therefore begin by briefly reviewing the key results of the seminal work, Ref~[\citen{beenakker_08}], based on the effective low-energy Hamiltonian $H_M$,
and then clarify the motivation for reexamining the problem from the perspective of pair correlations.
Within the analysis based on $H_M$, the scattering coefficients at low energies $E \ll E_M$ satisfy~\cite{beenakker_08}
\begin{align}
|s^{he}_{aa}| \ll |s^{he}_{ab}|=|s^{ee}_{ab}|,
\label{eq:scattering_heff}
\end{align}
where $s^{ee}_{\alpha \beta}$ and $s^{he}_{\alpha \beta}$ ($s^{eh}_{\alpha \beta}$ and $s^{hh}_{\alpha \beta}$)
denote the scattering coefficients from an electron (hole) in lead $\beta$ to an electron and a hole in lead $\alpha$, respectively.
Equation~(\ref{eq:scattering_heff}) highlights two important features.
First, inter-lead scattering dominates over local Andreev reflection.
Second, electron--electron tunneling (elastic cotunneling) and electron--hole tunneling (crossed Andreev reflection) occur with equal probability.
These properties give rise to an anomaly in the zero-frequency noise power, defined as
\begin{align}
P_{\alpha \beta} = \int_{-\infty}^{\infty} \overline{\delta I_{\alpha}(0) \delta I_{\beta}(t)} dt
\end{align}
where $\delta I_{\alpha}(t) = I_{\alpha}(t) - I_{\alpha}$ denotes the current fluctuations at time $t$,
and $I_{\alpha}$ is the time-averaged current in the lead $\alpha$.
In particular, for low bias voltages $eV \ll E_M$, the zero-frequency noise power satisfies~\cite{beenakker_08}
\begin{align}
2P_{ab}=P_{aa}+P_{bb},
\label{eq:max_cross_correlation}
\end{align}
indicating a maximal positive cross correlation, which is widely regarded as an experimental signature of Majorana nonlocality.
It is important to note, however, that this anomaly arises in the regime $eV \ll E_M$, whereas the nonlocal fermionic mode appears at $E=E_M$.
This mismatch of energy scales suggests that the unusual transport properties of a finite topological superconductor
cannot be attributed solely to the presence of the nonlocal fermionic mode.
In light of this, we reexamine the transport properties from the perspective of pair correlations to gain further insight into the origin of the maximal positive cross correlation.
%----------------------------------------------------------------------------
%----------------------------------------------------------------------------

To compute the scattering coefficients, we model the normal leads and the interfaces by
\begin{align}
\begin{split}
&H_{a}=-t \sum_{j <0}\left(a_{j+1}^{\dagger} a_j + \mathrm{h.c.}\right),\qquad
H_{b}=-t \sum_{j > L}\left(a_{j+1}^{\dagger} a_j + \mathrm{h.c.}\right),\\
&H_i=-t_i \left(c_{1}^{\dagger} a_0 + \mathrm{h.c.}\right)
-t_i \left(a_{L+1}^{\dagger} c_{L} + \mathrm{h.c.}\right),
\label{eq:normal_leads}
\end{split}
\end{align}
where the topological superconductor occupies the region $1\leq j \leq L$.
Here, $a_j$ ($a_j^\dagger$) annihilates (creates) an electron in the normal leads,
and $t_i$ denotes the hopping amplitude across the interfaces.
In the wide-band limit $E \ll t$, the scattering matrix is given by~\cite{ando_91}
\begin{align}
\begin{split}
&\check{S}=1-\frac{2i}{t}\check{T}_i^{\dagger}
\left(\check{G}^{-1}_s+\frac{i}{t}\check{T}_i\check{T}_i^{\dagger}\right)^{-1}\!\check{T}_i,\\
&\check{G}_s=
\left(\begin{array}{cc} \hat{G}(1,1) & \hat{G}(1,L) \\ \hat{G}(L,1) & \hat{G}(L,L) \end{array} \right), \qquad
\check{T}_i=
\left(\begin{array}{cc} -t_i\hat{\tau}_3 & 0 \\ 0 & -t_i\hat{\tau}_3 \end{array} \right),
\label{eq:scattering}
\end{split}
\end{align}
where
\begin{align}
\check{S}=\left(\begin{array}{cc}\hat{s}_{aa} & \hat{s}_{ab} \\ \hat{s}_{ba} & \hat{s}_{bb} \end{array} \right), \quad
\hat{s}_{\alpha \beta} = \left(\begin{array}{cc}s^{ee}_{\alpha \beta} & s^{eh}_{\alpha \beta} \\ s^{he}_{\alpha \beta} & s^{hh}_{\alpha \beta} \end{array} \right).
\end{align}
The retarded Green's function $\hat{G}(j,j^{\prime})$ is obtained from the Matsubara Green's function by analytic continuation,
\begin{align}
\hat{\mathcal{G}}(j,j') \xrightarrow{i\omega \to E + i0^+} \hat{G}(j,j').
\end{align}
A detailed derivation of Eq.~(\ref{eq:scattering}) is provided in Appendix~\ref{sec:app_scatt_green}.
In Nambu space, the retarded Green's function is written as
\begin{align}
\hat{G}(j,j')=\left( \begin{array}{cc} G(j,j') & F(j,j') \\ \underline{F}(j,j') & \underline{G}(j,j') \end{array} \right).
\end{align}
We consider the low-energy and weak-coupling regime
$E, t_i \ll E_M \ll \Delta_{k_F} \ll t$, for system lengths satisfying
$1 \ll \xi_0 \ll L$.
In this regime, the scattering matrix reduces to
\begin{align}
\check{S} \sim 1 -  \frac{2i}{t} \check{T}_i^{\dagger} \check{G}_s \check{T}_i.
\label{eq:scattering_eff}
\end{align}
The scattering coefficients are given by
\begin{align}
\begin{split}
&s^{he}_{aa} = \frac{2it_i^2}{t} \underline{F}(1,1)=\Gamma \frac{E}{E^2 - E_M^2} \sim 0,\\
&s^{he}_{ab} = \frac{2it_i^2}{t} \underline{F}(1,L)=\Gamma \frac{E_M}{E^2 - E_M^2} \sim -\frac{\Gamma}{E_M},\\
&s^{ee}_{ab} = -\frac{2it_i^2}{t} G(1,L)=-i\Gamma \frac{E_M}{E^2 - E_M^2} \sim i\frac{\Gamma}{E_M},
\label{eq:scattering_explicit}
\end{split}
\end{align}
where $\Gamma=2t_i^2C/t$.
The derivation of Eq.~(\ref{eq:scattering_eff}) is also provided in Appendix~\ref{sec:app_scatt_green}.
These results show that the key features of the scattering coefficients in Eq.~(\ref{eq:scattering_heff})
arise directly from the unusual properties of the Green's functions discussed in Sec.~\ref{sec:fundamental_properties}:
\begin{align}
|\underline{F}(1,1)| \ll |\underline{F}(1,L)|=|G(1,L)|.
\end{align}
In particular, inter-lead scattering dominates over local Andreev reflection due to the pronounced nonlocality of the Green's functions,
as demonstrated in Eqs.~(\ref{eq:prop2}) and (\ref{eq:prop2-2}).
Moreover, the equal amplitudes of elastic cotunneling and crossed Andreev reflection follow directly from the equivalence of the normal and anomalous Green's functions,
as shown in Eq.~(\ref{eq:prop1}).
From the perspective of pair correlations, the maximal positive cross correlation in Eq.~(\ref{eq:max_cross_correlation})
can thus be seen as a direct manifestation of the unconventional nonlocal Cooper pairs intrinsic to finite topological superconductors.

%****************************************************************************************************************
\section{Discussion and Summary}
%****************************************************************************************************************
In the present study, we focus on the Green's function in the low-temperature regime, $T \ll E_M$.
For much longer topological superconductors, however, the opposite limit, $E_M \ll T$, may be realized.
In this regime, the nonlocal Green's functions vanish due to equal contributions from the $E=+E_M$ and $E=-E_M$ states,
i.e., $\mathcal{G}(1,L) \propto \mathcal{F}(1,L) \propto \langle P_f \rangle = 0$ for $E_M/\omega \to 0$.
It has previously been pointed out that,
when the fermion parity is fixed, the end-to-end correlation function can remain finite even in the thermodynamic limit, $L \to \infty$~\cite{sato_16,zhou_18}.
Realizing such a fixed-parity situation generally requires additional ingredients, such as a charging energy~\cite{fu_10}.
Analyzing Green's functions in the presence of such interactions remains an important direction for future work.
In addition, establishing a direct connection between the Green's functions obtained here and experimentally accessible parity measurements is an important open problem~\cite{sarma_08}.
While we focus on systems hosting topologically protected Majorana end modes,
our theory can likely be extended to minimal Kitaev chains realized in quantum-dot--superconductor hybrid systems
hosting so-called poor man's Majorana modes~\cite{dassarma_12,leijnse_12,vanloo_26,alvarado_26}.
Investigating the nonlocal Green's functions in such systems would also be an intriguing direction for future work~\cite{cayao_24}.

In conclusion, we have examined the emergence of unconventional nonlocal Cooper pairs in finite topological superconductors,
characterized by two distinctive features of the Gor'kov Green's functions:
the equivalence between their normal and anomalous components, and their pronounced nonlocality.
We have shown that these nonlocal Cooper pairs are intrinsically linked to fermion parity and to the nonlocal transport properties of finite topological superconductors.
On the basis of the pair-correlation picture, our results provide further insight into Majorana nonlocality, a key concept in topological quantum computation.

\begin{acknowledgments}
We thank Y. Kawaguchi and Y. Tanaka for the fruitful discussion.
Y.N. is supported by JST SPRING (Grant No. JPMJSP2125) and thanks the THERS Make New Standards Program for the Next Generation Researchers.
S.I. is supported by a Grant-in-Aid for Early-Career Scientists (JSPS KAKENHI Grant No. JP24K17010). 
Y.A. is supported by a Grant-in-Aid for Scientific Research (JSPS KAKENHI Grant No. JP26K0692).
\end{acknowledgments}

\section*{Data availability}
The data that support the findings of this article are openly available \cite{data}.

\appendix
\section{Detailed derivations of  Gor'kov Green's functions}
\subsection{Bulk system}\label{sec:app_bulk}
In this section, we compute the Green's function for a bulk Kitaev chain.
Under periodic boundary conditions, the BdG Hamiltonian in Eq.~(\ref{eq:matrix_bdg}) is diagonalized as
\begin{align}
\sum_{j'}\hat{H}(j,j') \hat{U}_k(j')=\hat{U}_k(j') \left( \begin{array}{cc} E_k & 0 \\ 0 & -E_k \end{array} \right),
\end{align}
where
\begin{align}
\begin{split}
&\hat{U}_k (j) = \left( \begin{array}{cc} u_k & -v_k \\ v_k & u_k \end{array} \right) \frac{e^{ikj}}{\sqrt{L'}},\\
&u_k=\frac{E_k+\xi_k}{\sqrt{2E_k(E_k+\xi_k)}}, \quad
v_k=\frac{\Delta_k}{\sqrt{2E_k(E_k+\xi_k)}}, \\
&E_k = \sqrt{\xi_k^2 + \Delta_k^2},\quad
\xi_k=-2t \cos k -\mu, \quad \Delta_k = \Delta \sin k,
\end{split}
\end{align}
and $L'$ denotes the system length under periodic boundary conditions.
The Gor'kov Green's function satisfying Eq.~(\ref{eq:bulk_gorkov}) can be then written as
\begin{align}
\hat{\mathcal{G}}_0(j,j') = \sum_k \hat{U}_k(j) \left( \begin{array}{cc} i\omega - E_k & 0 \\ 0 &  i \omega + E_k \end{array} \right)^{-1} \! \hat{U}^{\dagger}_k(j') 
=\frac{1}{L'} \sum_k \sum_{\eta=\pm} \frac{e^{ik(j-j')}}{i \omega - \eta E_k} F_{\eta} (k)
\end{align}
where
\begin{align}
F_{\pm}(k) = \frac{1}{2E_k}
\left( \begin{array}{cc} E_k \pm \xi_k & \pm \Delta_k \\ \pm \Delta_k &  E_k \mp \xi_k \end{array} \right).
\end{align}
Taking the limit $L' \to \infty$, we obtain the bulk Green's function,
\begin{align}
\hat{\mathcal{G}}_0(j,j') = \frac{1}{2\pi} \int^{\pi}_{-\pi} dk \sum_{\eta=\pm} \frac{e^{ik(j-j')}}{i \omega - \eta E_k} F_{\eta} (k).
\end{align}
For $\omega \geq 0$ ($\omega<0$), the integrand with $\eta=+$ ($-$) has poles at complex wave numbers satisfying
\begin{align}
\cos k = \cos k_F - i \eta \frac{\Omega_k}{2t},
\end{align}
where
\begin{align}
k_F = \arccos \left( -\frac{\mu}{2t} \right), \qquad
\Omega_k = \sqrt{\omega^2 + \Delta^2_k}.
\end{align}
In the following, we assume $\omega, \Delta \ll t$,
and retain only the leading-order terms in $\delta = \omega/t, \Delta/t$.
In this regime, the pole positions reduce to
\begin{align}
k = \pm \left( k_F + i \eta \kappa \right),
\end{align}
where
\begin{align}
\kappa = \frac{1}{\xi} = \frac{\Omega_{k_F}}{v_F}, \quad v_F = 2t \sin k_F.
\end{align}
Evaluating the complex integral by computing the residues at these poles, we finally obtain
\begin{align}
\begin{split}
&\hat{\mathcal{G}}_0(j,j')= -\frac{i}{v_F \Omega_{k_F}}
\left(
a_0 + a_1 \hat{\tau}_1 + a_3 \hat{\tau}_3 \right),\\
&a_0=\omega \cos\!\big(k_F |j-j'|\big) e^{-\kappa |j-j'|},\\
&a_1=\Delta_{k_F} \sin\!\big(k_F (j-j')\big) e^{-\kappa |j-j'|},\\
&a_3=i\Omega_{k_F}\sin \!\big(k_F |j-j'|\big) e^{-\kappa |j-j'|}.
\end{split}
\end{align}

\subsection{Finite system}\label{sec:app_finite}
In this section, we compute the Green's function for a finite Kitaev chain.
We begin with the Dyson equation,
\begin{align}
\hat{\mathcal{G}}_V(j,j')
&= \hat{\mathcal{G}}_0(j,j')
+\sum_{j_1,j_2}
\hat{\mathcal{G}}_0(j,j_1)\hat{V}(j_1,j_2)\hat{\mathcal{G}}_V(j_2,j'),
\end{align}
where
\begin{align}
\hat{V}(j,j')
= V\,\delta_{j,j'} \bigl(\delta_{j,0}+\delta_{j,L+1}\bigr) \hat{\tau}_3
\end{align}
describes boundary potentials at $j=0$ and $j=L+1$.
Using the relation
\begin{align}
\check{A}_V \begin{pmatrix}  \hat{\mathcal{G}}_V(0,j') \\  \hat{\mathcal{G}}_V(\tilde{L},j') \end{pmatrix}
= \begin{pmatrix}  \hat{\mathcal{G}}_0(0,j') \\  \hat{\mathcal{G}}_0(\tilde{L},j') \end{pmatrix}
\end{align}
with
\begin{align}
\check{A}_V = \begin{pmatrix}  1-\hat{\mathcal{G}}_0(0,0) V \hat{\tau}_3 & -\hat{\mathcal{G}}_0(0,\tilde{L}) V \hat{\tau}_3 \\
-\hat{\mathcal{G}}_0(\tilde{L}, 0) V \hat{\tau}_3 & 1-\hat{\mathcal{G}}_0(\tilde{L},\tilde{L}) V \hat{\tau}_3 \end{pmatrix},
\end{align}
and $\tilde{L}=L+1$,
we obtain the closed-form expression:

\begin{align}
\hat{\mathcal{G}}_V(j,j')
= \hat{\mathcal{G}}_0(j,j')
+ \left( \hat{\mathcal{G}}_0(j, 0), \hat{\mathcal{G}}_0(j, \tilde{L}) \right)
\begin{pmatrix} V \hat{\tau}_3 & 0 \\ 0 & V \hat{\tau}_3 \end{pmatrix} A_V^{-1}
\begin{pmatrix}  \hat{\mathcal{G}}_0(0,j') \\  \hat{\mathcal{G}}_0(\tilde{L},j') \end{pmatrix}.
\end{align}
Taking the hard-wall limit, we obtain the Green's function for the finite topological superconductor for $1 \leq j,j' \leq L$:
\begin{align}
\hat{\mathcal{G}}(j,j') = \lim_{V\to\infty} \hat{\mathcal{G}}_V(j,j')
\; = \; \hat{\mathcal{G}}_0(j,j')
- \left( \hat{\mathcal{G}}_0(j, 0), \hat{\mathcal{G}}_0(j, \tilde{L}) \right)
\begin{pmatrix}  \hat{\mathcal{G}}_0(0,0) & \hat{\mathcal{G}}_0(0,\tilde{L}) \\
\hat{\mathcal{G}}_0(\tilde{L}, 0) & \hat{\mathcal{G}}_0(\tilde{L},\tilde{L}) \end{pmatrix}^{-1}
\begin{pmatrix}  \hat{\mathcal{G}}_0(0,j') \\  \hat{\mathcal{G}}_0(\tilde{L},j') \end{pmatrix}.
\end{align}
After straightforward algebra, we finally obtain
\begin{align}
\hat{\mathcal{G}}(j,j')=- i D
\left(b_0 + b_1 \hat{\tau}_1 + b_2 \hat{\tau}_2 + b_3 \hat{\tau}_3 \right),
\label{eq:green_finite_full}
\end{align}
where
\begin{align}
D
&=
\frac{1}{v_F \Omega_{k_F}}
\frac{1}{
\omega^2
\bigl[
\cosh^2(\kappa \tilde{L})
-
\cos^2(k_F \tilde{L})
\bigr]
+
\Delta_{k_F}^2
\sin^2(k_F \tilde{L})
},
\end{align}

\begin{align}
b_0
&=
-\frac{\omega^3}{2}
\Big[
\cos\!\big(k_F (j+j')\big)
\sinh\!\big(\kappa (2\tilde{L}-j-j')\big)
+
\cos\!\big(k_F (2\tilde{L}-j-j')\big)
\sinh\!\big(\kappa (j+j')\big)
\Big]
\nonumber\\
&\quad
+ \omega^3
\Big[
\cos\!\big(k_F (j-j')\big)
\cosh(\kappa \tilde{L})
\sinh\!\big(\kappa (\tilde{L}-|j-j'|)\big)
\nonumber\\
&\qquad
+
\cos(k_F \tilde{L})
\cos\!\big(k_F (\tilde{L}-|j-j'|)\big)
\sinh\!\big(\kappa |j-j'|\big)
\Big]
\nonumber\\
&\quad
+ \omega \Delta_{k_F}^2
\Big[
\sin(k_F j)\sin(k_F j')
\sinh\!\big(\kappa (2\tilde{L}-j-j')\big)
\nonumber\\
&\qquad
+
\sin\!\big(k_F (\tilde{L}-j)\big)
\sin\!\big(k_F (\tilde{L}-j')\big)
\sinh\!\big(\kappa (j+j')\big)
\Big]
\nonumber\\
&\quad
- \omega \Delta_{k_F}^2
\sin(k_F \tilde{L})
\sin\!\big(k_F (\tilde{L}-|j-j'|)\big)
\sinh\!\big(\kappa |j-j'|\big),
\end{align}

\begin{align}
b_1
&=
-\frac{\omega^2 \Delta_{k_F}}{2}
\sin\!\big(k_F (j-j')\big)
\Big[
\sinh\!\big(\kappa (2\tilde{L}-j-j')\big)
+
\sinh\!\big(\kappa (j+j')\big)
\Big]
\nonumber\\
&\quad
+ \omega^2 \Delta_{k_F}
\sin\!\big(k_F (j-j')\big)
\sinh(\kappa \tilde{L})
\cosh\!\big(\kappa (\tilde{L}-|j-j'|)\big)
\nonumber\\
&\quad
+ \Delta_{k_F} \Omega_{k_F}^2
\sin(k_F \tilde{L})
\, f(j,j')
\sinh\!\big(\kappa (j-j')\big),
\end{align}

\begin{align}
b_2
&=
\omega \Delta_{k_F} \Omega_{k_F}
\Big[
\sin\!\big(k_F(\tilde{L}-j)\big)
\sin\!\big(k_F(\tilde{L}-j')\big)
\nonumber\\
&\qquad\times
\cosh\!\big(\kappa (j+j')\big)
-
\sin(k_F j)\sin(k_F j') \cosh\!\big(\kappa (2\tilde{L}-j-j')\big)
\Big]
\nonumber\\
&\quad
+\frac{\omega \Delta_{k_F} \Omega_{k_F}}{2}
\Big[
\cos\!\big(k_F (2\tilde{L}-j-j')\big)
-
\cos\!\big(k_F (j+j')\big)
\Big] \cosh\!\big(\kappa (j-j')\big),
\end{align}

\begin{align}
b_3
&=
- i \frac{\omega^2 \Omega_{k_F}}{2}
\Big[
\sin\!\big(k_F (j+j')\big)\cosh\!\big(\kappa (2\tilde{L}-j-j')\big)
\nonumber\\
&\qquad
+
\sin\!\big(k_F (2\tilde{L}-j-j')\big)
\cosh\!\big(\kappa (j+j')\big)
\Big]
\nonumber\\
&\quad
+ i \omega \Omega_{k_F}
\Big[
\sin\!\big(k_F |j-j'| \big)
\cosh(\kappa \tilde{L})
\cosh\!\big(\kappa (\tilde{L}-|j-j'|)\big)
\nonumber\\
&\qquad
+
\cos(k_F \tilde{L})
\sin\!\big(k_F (\tilde{L}-|j-j'|)\big)
\cosh\!\big(\kappa (j-j')\big)
\Big]
\nonumber\\
&\quad
- i \Delta_{k_F}^2 \Omega_{k_F}
\sin(k_F \tilde{L})
\, f(j,j')
\cosh\!\big(\kappa (j-j')\big),
\end{align}
and
\begin{align}
f(j,j')= \begin{cases}
2 \sin \bigl(k_F(\tilde{L}-j)\bigr) \sin(k_F j')  & \text{for } j \geq j' \\
2  \sin(k_F j) \sin \bigl(k_F(\tilde{L}-j')\bigr)   & \text{for } j < j'
\end{cases}.
\end{align}

We next focus on the low-frequency regime, $\omega \ll \Delta_{k_F}$,
and consider a system length satisfying $1 \ll \xi_0 \ll L$,
where $\xi_0=v_F/\Delta_{k_F}$.
In what follows, we expand the Green's function up to leading order in 
\begin{align}
\delta = \frac{\omega}{\Delta_{k_F}},\;
\frac{1}{\xi_0},\;
e^{-L/\xi_0}.
\end{align}
In this regime, the prefactor of the Green's function can be approximated as
\begin{align}
D\sim \frac{1}{v_F \Delta_{k_F}}
\frac{4e^{-2L/\xi_0}}{\omega^2 +E_M^2},
\end{align}
where
\begin{align}
E_M = 2 \Delta_{k_F} \sin\!\big(k_F (L+1)\big) e^{-L/\xi_0}.
\end{align}
For $j=1$, the coefficients reduce to

\begin{align}
\begin{split}
b_0 \sim {}& \frac{\omega \Delta_{k_F}^2}{2} \sin(k_F) \sin(k_Fj')e^{(2L-j')/\xi_0},\\
b_1 \sim {}& - \Delta_{k_F}^3 \sin \!\big(k_F (L+1)\big) \sin (k_F) \sin(k_F (L+1-j'))e^{j'/\xi_0},\\
b_2 \sim {}& - \frac{\omega \Delta_{k_F}^2}{2} \sin(k_F) \sin(k_Fj')e^{(2L-j')/\xi_0},\\
b_3 \sim {}& - i \Delta_{k_F}^3 \sin \!\big(k_F (L+1)\big) \sin (k_F) \sin(k_F (L+1-j'))e^{j'/\xi_0}.
\end{split}
\end{align}
Similarly, the coefficients for $j'=L$ reduce to
\begin{align}
\begin{split}
b_0 \sim {}& \frac{\omega \Delta_{k_F}^2}{2} \sin(k_F (L+1-j)) \sin(k_F) e^{(L+j)/\xi_0},\\
b_1 \sim {}& \Delta_{k_F}^3 \sin \!\big(k_F (L+1)\big) \sin(k_Fj) \sin (k_F) e^{(L-j)/\xi_0},\\
b_2 \sim {}& \frac{\omega \Delta_{k_F}^2}{2}  \sin(k_F (L+1-j)) \sin(k_F) e^{(L+j)/\xi_0},\\
b_3 \sim {}& - i \Delta_{k_F}^3 \sin \!\big(k_F (L+1)\big)\sin(k_Fj) \sin (k_F) e^{(L-j)/\xi_0}.
\end{split}
\end{align}
Using these coefficients, we finally obtain
\begin{align}
\begin{split}
\hat{\mathcal{G}}(1,j')=&-i \frac{\omega}{\omega^2 + E_M^2} \frac{2}{\xi_0} \sin (k_F) \sin(k_F j')
\left(1 - \hat{\tau}_2\right) e^{-j'/\xi_0}\\
&-i \frac{E_M}{\omega^2 + E_M^2} \frac{2}{\xi_0} \sin (k_F) \sin(k_F (L+1-j'))
\left(-\hat{\tau}_1 - i\hat{\tau}_3\right) e^{-(L-j')/\xi_0},
\end{split}
\end{align}
and
\begin{align}
\begin{split}
\hat{\mathcal{G}}(j,L)=&-i \frac{\omega}{\omega^2 + E_M^2}  \frac{2}{\xi_0} \sin(k_F (L+1-j)) \sin (k_F)
\left(1 + \hat{\tau}_2\right) e^{-(L-j)/\xi_0}\\
&-i \frac{E_M}{\omega^2 + E_M^2}\frac{2}{\xi_0} \sin (k_F j) \sin(k_F)
\left(\hat{\tau}_1 - i\hat{\tau}_3\right) e^{-j/\xi_0}.
\end{split}
\end{align}

\section{Detailed derivation of the scattering matrix}\label{sec:app_scatt_green}
In this section, we compute the scattering matrix of a topological superconductor connected to two normal leads.
The normal leads and the junction interfaces are described by the Hamiltonians given in Eq.~(\ref{eq:normal_leads}).
We start from a general expression for the scattering matrix~\cite{ando_91},
\begin{align}
\check{S}=\check{G} \check{Q} -1,
\end{align}
where
\begin{align}
\begin{split}
&\check{G}=\check{G}_n+\check{G}_n\check{T}_i^{\dagger}
\left(\check{G}^{-1}_s-\check{T}_i\check{G}_n \check{T}_i^{\dagger}\right)^{-1}\!\check{T}_i \check{G}_n,\\
&\check{G}_n=
\left(\begin{array}{cc} \hat{G}_n & 0 \\ 0 & \hat{G}_n \end{array} \right), \quad
\hat{G}_n=
\left(\begin{array}{cc} g_n & 0 \\ 0 & \underline{g}_n \end{array} \right),\\
&\check{G}_s=
\left(\begin{array}{cc} \hat{G}(1,1) & \hat{G}(1,L) \\ \hat{G}(L,1) & \hat{G}(L,L) \end{array} \right), \quad
\check{T}_i=
\left(\begin{array}{cc} -t_i\hat{\tau}_3 & 0 \\ 0 & -t_i\hat{\tau}_3 \end{array} \right),\\
\end{split}
\end{align}
and
\begin{align}
\check{Q}=
\left(\begin{array}{cc} \hat{Q} & 0 \\ 0 & \hat{Q} \end{array} \right), \quad
\hat{Q}=
\left(\begin{array}{cc} q & 0 \\ 0 & \underline{q} \end{array} \right),
\end{align}
with
\begin{align}
\begin{split}
g_n = (E+t\lambda_+)^{-1}, \quad 
q = t (\lambda_+-\lambda_-),\quad
\lambda_{\pm}=\frac{A \pm i \sqrt{4-A^2}}{2}, \quad A = - \frac{E}{t},
\end{split}
\end{align}
and
\begin{align}
\begin{split}
\underline{g}_n = (E-t\underline{\lambda}_+)^{-1}, \quad 
\underline{q} = -t (\underline{\lambda}_+-\underline{\lambda}_-),\quad
\underline{\lambda}_{\pm}=\frac{\underline{A} \mp i \sqrt{4-\underline{A}^2}}{2}, \quad \underline{A} = \frac{E}{t}.
\end{split}
\end{align}
The retarded Green's function $\hat{G}(j,j^{\prime})$ is obtained from the Matsubara Green's function by analytic continuation,
\begin{align}
\hat{\mathcal{G}}(j,j') \xrightarrow{i\omega \to E + i0^+} \hat{G}(j,j').
\end{align}
For the wide-band limit $E/t \to 0$, the scattering matrix reduces to
\begin{align}
\check{S}=1-\frac{2i}{t}\check{T}_i^{\dagger}
\left(\check{G}^{-1}_s+\frac{i}{t}\check{T}_i\check{T}_i^{\dagger}\right)^{-1}\!\check{T}_i.
\end{align}
In the regime, $E \ll \Delta_{k_F}$,
and $1 \ll \xi_0 \ll L$,
the retarded Green's functions are given by
\begin{align}
\begin{split}
\hat{G}(1,1)
= C \frac{E}{E^2 - E_M^2}
\left(1 - \hat{\tau}_2\right), \qquad
\hat{G}(L,L)
= C \frac{E}{E^2 - E_M^2}
\left(1 + \hat{\tau}_2\right),
\end{split}
\end{align}
and
\begin{align}
\begin{split}
\hat{G}(1,L)
= i C \frac{E_M}{E^2 - E_M^2}
\left(-\hat{\tau}_1 - i\hat{\tau}_3\right), \qquad
\hat{G}(L,1)
= i C \frac{E_M}{E^2 - E_M^2}
\left(\hat{\tau}_1 - i\hat{\tau}_3\right).
\end{split}
\end{align}
Further assuming the low-energy limit, $E \ll E_M \ll \Delta_{k_F}$,
we obtain
\begin{align}
\begin{split}
G(1,1) \propto \frac{E}{E_M^2}\frac{\Delta_{k_F}}{t} \sim 0,\qquad
G(L,L) \propto \frac{E}{E_M^2}\frac{\Delta_{k_F}}{t} \sim 0,
\end{split}
\end{align}
and
\begin{align}
\begin{split}
G(1,L) \propto \frac{1}{E_M}\frac{\Delta_{k_F}}{t},\qquad
G(L,L) \propto \frac{1}{E_M}\frac{\Delta_{k_F}}{t}.
\end{split}
\end{align}
Thus, in the tunneling regime satisfying
\begin{align}
\frac{t_i^2}{t} \ll \left( \frac{1}{E_M}\frac{\Delta_{k_F}}{t} \right)^{-1}
\; \Rightarrow \;
\left(\frac{t_i}{t}\right)^2 \ll \frac{E_M}{\Delta_{k_F}},
\end{align}
we find
\begin{align}
\left(\check{G}^{-1}_s+\frac{i}{t}\check{T}_i\check{T}_i^{\dagger}\right)^{-1} \sim \check{G}_s.
\end{align}
Consequently, the scattering matrix in this regime reduces to
\begin{align}
\check{S} \sim 1 -  \frac{2i}{t} \check{T}_i^{\dagger} \check{G}_s \check{T}_i.
\end{align}


\begin{thebibliography}{}
%
%----- Proximity Effect -----
%
\bibitem{andreev_64}A.~F.~Andreev, 
Thermal Conductivity of the Intermediate State of Superconductors,
Sov. Phys. JETP \textbf{19}, 1228 (1964).
\bibitem{beenakker_97}C.~W.~J.~Beenakker,
Random-matrix theory of quantum transport
\href{https://journals.aps.org/rmp/abstract/10.1103/RevModPhys.69.731}{Rev. Mod. Phys. \textbf{69}, 731 (1997)}.
%
\bibitem{courtois_00}B.~Pannetier and H.~Courtois,
Andreev Reflection and Proximity effect,
\href{https://link.springer.com/article/10.1023/A:1004635226825}{J. Low. Tmp. Phys. \textbf{118}, 599 (2000)}.
%
\bibitem{gennes_64}P.~G.~de~Gennes,
Boundary Effects in Superconductors,
\href{https://journals.aps.org/rmp/abstract/10.1103/RevModPhys.36.225}{Rev. Mod. Phys. \textbf{36}, 225 (1964)}.
%
\bibitem{eilenberger_68} G.~Eilenberger,
Transformation of Gorkov’s Equation for Type II Superconductors into Transport-Like Equations,
\href{https://link.springer.com/article/10.1007/BF01379803}{Zeitschrift f\"ur Physik \textbf{214}, 195--213 (1968)}.
%
\bibitem{usadel_70} K.~D.~Usadel,
Generalized Diffusion Equation for Superconducting Alloys,
\href{https://journals.aps.org/prl/abstract/10.1103/PhysRevLett.25.507}{Phys. Rev. Lett. \textbf{25}, 507 (1970)}.
%
%----- Majorana end modes -----
%
\bibitem{green_00} N.~Read and D.~Green,
Paired states of fermions in two dimensions with breaking of parity and time-reversal symmetries and the fractional quantum Hall effect,
\href{https://journals.aps.org/prb/abstract/10.1103/PhysRevB.61.10267}{Phys. Rev. B \textbf{61}, 10267 (2000)}.
%
\bibitem{kitaev_01} A.~Y.~Kitaev,
Unpaired Majorana fermions in quantum wires,
\href{https://iopscience.iop.org/article/10.1070/1063-7869/44/10S/S29}{Phys. Usp. \textbf{44}, 131 (2001)}.
%
\bibitem{kane_10} M.~Z.~Hasan and C.~L.~Kane,
Colloquium: Topological insulators,
\href{https://journals.aps.org/rmp/abstract/10.1103/RevModPhys.82.3045}{Rev. Mod. Phys. \textbf{82}, 3045 (2010)}.
%
\bibitem{zhang_11}  X.-L.~Qi  and  S.-C.~Zhang,
Topological insulators and superconductors,
\href{https://journals.aps.org/rmp/abstract/10.1103/RevModPhys.83.1057}{Rev. Mod. Phys. \textbf{83}, 1057 (2011)}.
%
\bibitem{sato_16} M. Sato and S. Fujimoto,
Majorana Fermions and Topology in Superconductors,
\href{https://journals.jps.jp/doi/10.7566/JPSJ.85.072001}{J. Phys. Soc. Jpn. \textbf{85}, 072001 (2016)}.
%
\bibitem{strinati_26} M.~C.~Strinati and G.~C.~Strinati,
Analysis of the singular band structure occurring in one-dimensional topological normal and superfluid fermionic systems: A pedagogical description,
\href{https://pubs.aip.org/avs/aqs/article/8/1/016201/3380244/Analysis-of-the-singular-band-structure-occurring}{AVS Quantum Sci. \textbf{8}, 016201 (2026)}.
%
%----- Odd-frequency pairing -----
%
\bibitem{berezinskii_74} V.~L.~Berezinskii,
New model of the anisotropic phase of superfluid He3,
JETP Lett. \textbf{20}, 287 (1974).
%
\bibitem{efetov_05} F.~S.~Bergeret, A.~F.~Volkov, and K.~B.~Efetov,
Odd triplet superconductivity and related phenomena in superconductor-ferromagnet structures,
\href{https://journals.aps.org/rmp/abstract/10.1103/RevModPhys.77.1321}{Rev. Mod. Phys. \textbf{77}, 1321 (2005)}.
%
\bibitem{tanaka_07} Y.~Tanaka and A.~A.~Golubov,
Theory of the Proximity Effect in Junctions with Unconventional Superconductors,
\href{https://journals.aps.org/prl/abstract/10.1103/PhysRevLett.98.037003}{Phys. Rev. Lett. \textbf{98}, 037003 (2007)}.
%
\bibitem{linder_19} J.~Linder and A.~V.~Balatsky,
Odd-frequency superconductivity,
\href{https://journals.aps.org/rmp/abstract/10.1103/RevModPhys.91.045005}{Rev. Mod. Phys. \textbf{91}, 045005 (2019)}.
%
%----- Majorana and Odd-frequency -----
%
\bibitem{nagaosa_12} Y.~Tanaka, M.~Sato, and N.~Nagaosa,
Symmetry and Topology in Superconductors --Odd-Frequency Pairing and Edge States--,
\href{https://journals.jps.jp/doi/abs/10.1143/JPSJ.81.011013}{J. Phys. Soc. Jpn. \textbf{81}, 011013 (2012)}.
%
\bibitem{asano_13} Y.~Asano and Y.~Tanaka,
Majorana fermions and odd-frequency Cooper pairs in a normal-metal nanowire proximity-coupled to a topological superconductor,
\href{https://journals.aps.org/prb/abstract/10.1103/PhysRevB.87.104513}{Phys. Rev. B \textbf{87}, 104513 (2013)}. 
%
\bibitem{cayao_20} J.~Cayao, C.~Triola, and A.~M.~Black-Schaffer,
Odd-frequency superconducting pairing in one-dimensional systems,
\href{https://link.springer.com/article/10.1140/epjst/e2019-900168-0}{Eur. Phys. J. Special Topics \textbf{229}, 545 (2020)}.
%
\bibitem{gurarie_11}V.~Gurarie,
Single-particle Green’s functions and interacting topological insulators,
\href{https://journals.aps.org/prb/abstract/10.1103/PhysRevB.83.085426}{Phys. Rev. B \textbf{83}, 085426 (2011)}.
%
\bibitem{tamura_19} S.~Tamura, S.~Hoshino, and Y.~Tanaka,
Odd-frequency pairs in chiral symmetric systems: Spectral bulk-boundary correspondence and topological criticality
\href{https://journals.aps.org/prb/abstract/10.1103/PhysRevB.99.184512}{Phys. Rev. B \textbf{99}, 184512 (2019)}.
%
\bibitem{daido_19} A.~Daido and Y.~Yanase,
Chirality polarizations and spectral bulk-boundary correspondence,
\href{https://journals.aps.org/prb/abstract/10.1103/PhysRevB.100.174512}{Phys. Rev. B \textbf{100}, 174512 (2019)}.
%
%----- Anomalous transport -----
\bibitem{tanaka_04} Y. Tanaka and S. Kashiwaya,
Anomalous charge transport in triplet superconductor junctions,
\href{https://journals.aps.org/prb/abstract/10.1103/PhysRevB.70.012507}{Phys. Rev. B \textbf{70}, 012507 (2004)}.
%
\bibitem{tanaka_05(1)} Y. Tanaka, S. Kashiwaya, and T. Yokoyama,
Theory of enhanced proximity effect by midgap Andreev resonant state in diffusive normal-metal/triplet superconductor junctions,
\href{https://journals.aps.org/prb/abstract/10.1103/PhysRevB.71.094513}{Phys. Rev. B \textbf{71}, 094513 (2005)}.
%
\bibitem{law_09}K. T. Law, P. A. Lee, and T. K. Ng,
Majorana Fermion Induced Resonant Andreev Reflection,
\href{https://journals.aps.org/prl/abstract/10.1103/PhysRevLett.103.237001}{Phys. Rev. Lett. \textbf{103}, 237001 (2009)}.
%
\bibitem{ikegaya_15} S. Ikegaya, Y. Asano, and Y. Tanaka,
Anomalous proximity effect and theoretical design for its realization,
\href{https://journals.aps.org/prb/abstract/10.1103/PhysRevB.91.174511}{Phys. Rev. B \textbf{91}, 174511 (2015)}.
%
\bibitem{ikegaya_16(1)} S. Ikegaya, S.-I. Suzuki, Y. Tanaka, and Y. Asano,
Quantization of conductance minimum and index theorem,
\href{https://journals.aps.org/prb/abstract/10.1103/PhysRevB.94.054512}{Phys. Rev. B \textbf{94}, 054512 (2016)}.
%
%----- Paramagnetic response ----
%
\bibitem{asano_11} Y.~Asano, A.~A.~Golubov, Y.~V. Fominov, and Y.~Tanaka,
Unconventional Surface Impedance of a Normal-Metal Film Covering a Spin-Triplet Superconductor Due to Odd-Frequency Cooper Pairs,
\href{https://journals.aps.org/prl/abstract/10.1103/PhysRevLett.107.087001}{Phys. Rev. Lett. \textbf{107}, 087001 (2011)}.
%
\bibitem{higashitani_14} S.~Higashitani, H.~Takeuchi, S.~Matsuo, Y.~Nagato, and K.~Nagai,
Magnetic Response of Odd-Frequency $s$-Wave Cooper Pairs in a Superfluid Proximity System,
\href{https://journals.aps.org/prl/abstract/10.1103/PhysRevLett.110.175301}{Phys. Rev. Lett. \textbf{110}, 175301 (2014)}.
%
\bibitem{asano_14} S.-I.~Suzuki and Y.~Asano,
Paramagnetic instability of small topological superconductors,
\href{https://journals.aps.org/prb/abstract/10.1103/PhysRevB.89.184508}{Phys. Rev. B \textbf{89}, 184508 (2014)}.
%
\bibitem{lutchyn_17} S.-P.~Lee, R.~M.~Lutchyn, and J.~Maciejko,
Odd-frequency superconductivity in a nanowire coupled to Majorana zero modes,
\href{https://journals.aps.org/prb/abstract/10.1103/PhysRevB.95.184506}{Phys. Rev. B \textbf{95}, 184506 (2017)}.
%
% ----- Topological quantum computation -----
%
\bibitem{ivanov_01} D.~A.~Ivanov,
Non-Abelian Statistics of Half-Quantum Vortices in $p$-Wave Superconductors,
\href{https://journals.aps.org/prl/abstract/10.1103/PhysRevLett.86.268}{Phys. Rev. Lett. \textbf{86}, 268 (2001)}.
%
\bibitem{kitaev_03} A.~Y.~Kitaev,
Fault-tolerant quantum computation by anyons,
\href{https://www.sciencedirect.com/science/article/abs/pii/S0003491602000180?via\%3Dihub}{Ann. Phys. \textbf{303}, 2-30 (2003)}.
%
\bibitem{sarma_08} C.~Nayak, S.~H.~Simon, A.~Stern, M.~Freedman, and S.~Das~Sarma,
Non-Abelian anyons and topological quantum computation,
\href{https://journals.aps.org/rmp/abstract/10.1103/RevModPhys.80.1083}{Rev. Mod. Phys. \textbf{80}, 1083(2008)}.
%
\bibitem{fisher_11} J.~Alicea, Y.~Oreg, G.~Refael, F.~v.~Oppen, and M.~P.~A.~Fisher,
Non-Abelian statistics and topological quantum information processing in 1D wire networks,
\href{https://www.nature.com/articles/nphys1915}{Nat. Phys. \textbf{7}, 412-417(2011)}.
%
\bibitem{alicea_16} D.~Aasen, M.~Hell, R.~V.~Mishmash, A.~Higginbotham, J.~Danon, M.~Leijnse, T.~S.~Jespersen, J.~A.~Folk, C.~M.~Marcus, K.~Flensberg, and J.~Alicea,
Milestones Toward Majorana-Based Quantum Computing,
\href{https://journals.aps.org/prx/abstract/10.1103/PhysRevX.6.031016}{Phys. Rev. X \textbf{6}, 031016 (2016)}.
%
% ----- Nonlocal transport -----
%
\bibitem{demler_07} C.~J.~Bolech and E.~Demler,
Observing Majorana bound States in $p$-Wave Superconductors Using Noise Measurements in Tunneling Experiments,
\href{https://journals.aps.org/prl/abstract/10.1103/PhysRevLett.98.237002}{Phys. Rev. Lett. \textbf{98}, 237002 (2007)}.
%
\bibitem{tewari_08}S.~Tewari, C.~Zhang, S.~Das~Sarma, C.~Nayak, and D.-H.~Lee,
Testable Signatures of Quantum Nonlocality in a Two-Dimensional Chiral $p$-Wave Superconductor,
\href{https://journals.aps.org/prl/abstract/10.1103/PhysRevLett.100.027001}{Phys. Rev. Lett. \textbf{100}, 027001 (2008)}.
%
\bibitem{beenakker_08}J.~Nilsson, A.~R.~Akhmerov, and C.~W.~J.~Beenakker,
Splitting of a Cooper Pair by a Pair of Majorana Bound States,
\href{https://journals.aps.org/prl/abstract/10.1103/PhysRevLett.101.120403}{Phys. Rev. Lett. \textbf{101}, 120403 (2008)}.
%
\bibitem{fu_10} L.~Fu,
Electron Teleportation via Majorana Bound States in a Mesoscopic Superconductor,
\href{https://journals.aps.org/prl/abstract/10.1103/PhysRevLett.104.056402}{Phys. Rev. Lett. \textbf{104}, 056402 (2010)}.
%
\bibitem{egger_11}A.~Zazunov, A.~L.~Yeyati, and R.~Egger,
Coulomb blockade of Majorana-fermion-induced transport,
\href{https://journals.aps.org/prb/abstract/10.1103/PhysRevB.84.165440}{Phys. Rev. B \textbf{84}, 165440 (2011)}.
%
\bibitem{hu_13}Z.~Wang, X.-Y.~Hu, Q.-F.~Liang, and X.~Hu,
Detecting Majorana fermions by nonlocal entanglement between quantum dots,
\href{https://journals.aps.org/prb/abstract/10.1103/PhysRevB.87.214513}{Phys. Rev. B \textbf{87}, 214513 (2013)}.
%
\bibitem{law_13}J.~Liu1, F,-C.~Zhang, and K.~T.~Law,
Majorana fermion induced nonlocal current correlations in spin-orbit coupled superconducting wires,
\href{https://journals.aps.org/prb/abstract/10.1103/PhysRevB.88.064509}{Phys. Rev. B \textbf{88}, 064509 (2013)}.
%
\bibitem{rosenow_13}B.~Zocher and B.~Rosenow,
Modulation of Majorana-Induced Current Cross-Correlations by Quantum Dots,
\href{https://journals.aps.org/prl/abstract/10.1103/PhysRevLett.111.036802}{Phys. Rev. Lett. \textbf{111}, 036802 (2013)}.
%
\bibitem{tewari_15}J.~D.~Sau, B.~Swingle, and S.~Tewari,
Proposal to probe quantum nonlocality of Majorana fermions in tunneling experiments,
\href{https://journals.aps.org/prb/abstract/10.1103/PhysRevB.92.020511}{Phys. Rev. B \textbf{92}, 020511(R) (2015)}.
%
\bibitem{flensberg_20}J.~Danon, A.~B.~Hellenes, E.~B.~Hansen, L.~Casparis, A.~P.~Higginbotham, and K.~Flensberg,
Nonlocal Conductance Spectroscopy of Andreev Bound States: Symmetry Relations and BCS Charges
\href{https://journals.aps.org/prl/abstract/10.1103/PhysRevLett.124.036801}{Phys. Rev. Lett. \textbf{124}, 036801 (2020)}.
%
\bibitem{xu_20}X.-Q.~Li and L.~Xu,
Nonlocality of Majorana zero modes and teleportation: Self-consistent treatment based on the Bogoliubov–de Gennes equation,
\href{https://journals.aps.org/prb/abstract/10.1103/PhysRevB.101.205401}{Phys. Rev. B \href{101}, 205401 (2020)}.
%
\bibitem{fujimoto_23} M.~Sugeta, T.~Mizushima, and S.~Fujimoto,
Enhanced 2$\pi$-periodic Aharonov--Bohm Effect as a Signature of Majorana Zero Modes Probed by Nonlocal Measurements,
\href{https://journals.jps.jp/doi/10.7566/JPSJ.92.054701}{J. Phys. Soc. Jpn. \textbf{92}, 054701 (2023)}.
%
\bibitem{ikegaya_24}Y.~Nagae, A.~P.~Schnyder, Y.~Tanaka, Y.~Asano, and S.~Ikegaya,
Multilocational Majorana zero modes,
\href{https://journals.aps.org/prb/abstract/10.1103/PhysRevB.110.L041110}{Phys. Rev. B \textbf{110}, L041110 (2024)}.
%
%--Coupling energy----
%
\bibitem{stanescu_12}S.~Das~Sarma, J.~D.~Sau, and T.~D.~Stanescu,
Splitting of the zero-bias conductance peak as smoking gun evidence for the existence of the Majorana mode in a superconductor-semiconductor nanowire,
\href{https://journals.aps.org/prb/abstract/10.1103/PhysRevB.86.220506}{Phys. Rev. B \textbf{86}, 220506(R) (2012)}.
%
%----- End-to-end correlation functions -----
%
\bibitem{zhou_18} J.-J.~Mioa, H.-K.~Jin, F.-C.~Zhang, Y.~Zhou,
Majorana zero modes and long range edge correlation in interacting Kitaev chains: analytic solutions and density-matrixrenormalization-group study,
\href{https://www.nature.com/articles/s41598-017-17699-y}{Sci. Rep. \textbf{8}, 488 (2018)}.
%
%
\bibitem{tanaka_24}Y.~Tanaka, S.~Tamura, and J.~Cayao,
Theory of Majorana zero modes in unconventional superconductors,
\href{https://academic.oup.com/ptep/article/2024/8/08C105/7663572?login=true}{Prog. Theor. Exp. Phys. \textbf{2024}, 08C105 (2024).}
%
%----- Semiconductor--superconductor nanowire -----
%
\bibitem{sarma_10} R.~M.~Lutchyn, J.~D.~Sau, and S.~DasSarma,
Majorana Fermions and a Topological Phase Transition in Semiconductor-Superconductor Heterostructures,
\href{https://journals.aps.org/prl/abstract/10.1103/PhysRevLett.105.077001}{Phys. Rev. Lett. \textbf{105}, 077001 (2010)}.
%
\bibitem{oreg_10} Y.~Oreg, G.~Refael, and F.~von~Oppen,
Helical Liquids and Majorana Bound States in Quantum Wires,
\href{https://journals.aps.org/prl/abstract/10.1103/PhysRevLett.105.177002}{Phys. Rev. Lett. \textbf{105}, 177002 (2010)}.
%
\bibitem{kouwenhoven_12} V.~Mourik, K.~Zuo, S.~M.~Frolov, S.~R.~Plissard, E.~P.~A.~M. Bakkers, and L.~P.~Kouwenhoven,
Signatures of Majorana Fermions in Hybrid Superconductor-Semiconductor Nanowire Devices,
\href{https://science.sciencemag.org/content/336/6084/1003}{Science \textbf{336}, 1003-1007 (2012)}.
%
\bibitem{deng_12} M.~T.~Deng, C.~L.~Yu, G.~Y.~Huang, M.~Larsson, P.~Caroff, and H.~Q.~Xu,
Anomalous Zero-Bias Conductance Peak in a Nb--InSb Nanowire--Nb Hybrid Device,
\href{https://pubs.acs.org/doi/10.1021/nl303758w}{Nano Lett. \textbf{12}, 6414-6419 (2012)}.
% Magnetic atom chain on a superconductor
\bibitem{beenakker_11} T.-P.~Choy, J.~M.~Edge, A.~R.~Akhmerov, and C.~W.~J.~Beenakker,
Majorana fermions emerging from magnetic nanoparticles on a superconductor without spin-orbit coupling,
\href{https://journals.aps.org/prb/abstract/10.1103/PhysRevB.84.195442}{Phys. Rev. B \textbf{84}, 195442 (2011)}.
%
\bibitem{yazdani_13} S.~Nadj-Perge, I.~K.~Drozdov, B.~A.~Bernevig, and A.~Yazdani,
Proposal for realizing Majorana fermions in chains of magnetic atoms on a superconductor,
\href{https://journals.aps.org/prb/abstract/10.1103/PhysRevB.84.195442}{Phys. Rev. B \textbf{88}, 020407(R) (2013)}.
%
\bibitem{yazdani_14} S.~Nadj-Perge, I.~K.~Drozdov, J.~Li, H.~Chen, S.~Jeon, J.~Seo, A.~H.~MacDonald, B.~A.~Bernevig, and A.~Yazdani,
Observation of Majorana fermions in ferromagnetic atomic chains on a superconductor,
\href{https://science.sciencemag.org/content/346/6209/602}{Science \textbf{346}, 602-607 (2014)}.
%
\bibitem{yazdani_17} B.~E.~Feldman, M.~T.~Randeria, J.~Li, S.~Jeon, Y.~Xie, Z.~Wang, I.~K.~Drozdov,  B.~A.~Bernevig, and A.~Yazdani,
High-resolution studies of the Majorana atomic chain platform,
\href{https://www.nature.com/articles/nphys3947}{Nat. Phys. \textbf{13}, 286-291 (2017)}.
% Topological Josephson junctions
\bibitem{flensberg_17} M.~Hell, M.~Leijnse, and K.~Flensberg,
Two-Dimensional Platform for Networks of Majorana Bound States,
\href{https://journals.aps.org/prl/abstract/10.1103/PhysRevLett.118.107701}{Phys. Rev. Lett. \textbf{118}, 107701 (2017)}.
%
\bibitem{halperin_17} F.~Pientka, A.~Keselman, E.~Berg, A.~Yacoby, A.~Stern, and B.~I.~Halperin,
Topological Superconductivity in a Planar Josephson Junction,
\href{https://journals.aps.org/prx/abstract/10.1103/PhysRevX.7.021032}{Phys. Rev. X \textbf{7}, 021032 (2017)}.
%
\bibitem{nichele_19} A.~Fornieri, A.~M.~Whiticar, F.~Setiawan, E.~Portol\'{e}s, A.~C.~C.~Drachmann, A.~Keselman, S.~Gronin, C.~Thomas, T.~Wang, R.~Kallaher,
G.~C.~Gardner, E.~Berg, M.~J.~Manfra, A.~Stern, C.~M.~Marcus, and F.~Nichele,
Evidence of topological superconductivity in planar Josephson junctions,
\href{https://www.nature.com/articles/s41586-019-1068-8}{Nature \textbf{569}, 89-92 (2019)}.
%
\bibitem{yacoby_19} H.~Ren, F.~Pientka, S.~Hart, A.~T.~Pierce, M.~Kosowsky, L.~Lunczer, R.~Schlereth,
B.~Scharf, E.~M.~Hankiewicz, L.~W.~Molenkamp, B.~I.~Halperin, and A.~Yacoby,
Topological superconductivity in a phase-controlled Josephson junction,
\href{https://www.nature.com/articles/s41586-019-1148-9}{Nature \textbf{569}, 93-98 (2019)}.
%
\bibitem{ikegaya_22} D.~Oshima, S.~Ikegaya, A.~P.~Schnyder, and Y.~Tanaka,
Flat-band Majorana bound states in topological Josephson junctions,
\href{https://journals.aps.org/prresearch/abstract/10.1103/PhysRevResearch.4.L022051}{Phys. Rev. Research \textbf{4}, L022051 (2022)}.
%
% Recursive Green's function techniques
%
\bibitem{fisher_81} P.~A.~Lee and D.~S.~Fisher,
Anderson Localization in Two Dimensions,
\href{https://journals.aps.org/prl/abstract/10.1103/PhysRevLett.47.882}{Phys. Rev. Lett. \textbf{47}, 882 (1981)}.
\bibitem{ando_91} T.~Ando,
Quantum point contacts in magnetic fields,
\href{https://journals.aps.org/prb/abstract/10.1103/PhysRevB.44.8017}{Phys. Rev. B \textbf{44}, 8017 (1991)}.
%
% Poorman's Majorana
%
\bibitem{dassarma_12}J.~D.~Sau and S.~Das~Sarma,
Realizing a robust practical Majorana chain in a quantum-dot-superconductor linear array,
\href{https://www.nature.com/articles/ncomms1966}{Nat. Commun. \textbf{3}, 964 (2012)}.
%
\bibitem{leijnse_12}M.~Leijnse and K.~Flensberg,
Parity qubits and poor man's Majorana bound states in double quantum dots,
\href{https://journals.aps.org/prb/abstract/10.1103/PhysRevB.86.134528}{Phys. Rev. B \textbf{86}, 134528 (2012)}.
%
\bibitem{vanloo_26}N.~van~Loo, F.~Zatelli, G.~O.~Steffensen, B.~Roovers, G.~Wang, T.~V.~Caekenberghe, A.~Bordin,
D.~van~Driel, Y.~Zhang, W.~D.~Huisman, G.~Badawy, E.~P.~A.~M.~Bakkers, G.~P.~Mazur, R.~Aguado, and L.~P.~Kouwenhoven,
Single-shot parity readout of a minimal Kitaev chain,
\href{https://www.nature.com/articles/s41586-025-09927-7}{Nature \textbf{650}, 334--339 (2026)}.
%
\bibitem{alvarado_26}
M.~Alvarado, R.~S.~Souto, M.~J.~Calder\'on, and R.~Aguado,
Optimal Majoranas in Mesoscopic Kitaev Chains,
\href{https://arxiv.org/abs/2604.13945v1}{arXiv:2604.13945 (2026)}.
%
\bibitem{cayao_24}J.~Cayao,
Emergent pair symmetries in systems with poor man's Majorana modes,
\href{https://journals.aps.org/prb/abstract/10.1103/PhysRevB.110.125408}{Phys. Rev. B \textbf{110}, 125408 (2024)}.
%
%----- Data -----
%
\bibitem{data} \href{https://github.com/ikegayas/data_2604.25169}{https://github.com/ikegayas/data\_2604.25169}.
\end{thebibliography}
\end{document}